\documentclass[preprint,unsortedaddress]{revtex4-1}
\usepackage[pdftex]{graphicx}
\usepackage[english]{babel}
\usepackage{float}
\usepackage{amssymb}   
\usepackage{amsmath}   
\usepackage{epsfig}
\usepackage{dcolumn}   
\usepackage{bm}        

\begin{document}

\title{Using CIPSI nodes in diffusion Monte Carlo}

\author{Michel Caffarel}
\affiliation{Lab. Chimie et Physique Quantiques, CNRS-Universit\'e de Toulouse, France}
\author{Thomas Applencourt}
\affiliation{Lab. Chimie et Physique Quantiques, CNRS-Universit\'e de Toulouse, France}
\author{Emmanuel Giner}
\affiliation{Dipartimento di Scienze Chimiche e Farmaceutiche, Università degli Studi di Ferrara, Ferrara, Italy}
\author{Anthony Scemama}
\affiliation{Lab. Chimie et Physique Quantiques, CNRS-Universit\'e de Toulouse, France}


\begin{abstract}
Several aspects of the recently proposed DMC-CIPSI approach consisting in using selected 
Configuration Interaction (SCI) approaches such as CIPSI 
(Configuration Interaction using a Perturbative 
Selection done Iteratively) to build accurate nodes for diffusion Monte Carlo (DMC) 
calculations are presented and discussed. The main ideas 
are illustrated with a number of calculations for diatomics molecules and for the benchmark G1 set.
\end{abstract}

\maketitle

\section{Introduction}
\label{intro}
In recent years the present authors have reported a number of fixed-node DMC studies 
using trial wavefunctions whose determinantal part is built with the CIPSI approach.
\cite{giner_2013,giner_phd_2014,scemama_jcp_2014,giner_jcp_2015,caffarel_jcp_2016}
The purpose of this paper is to review the present situation, to clarify 
some important aspects of DMC-CIPSI, and to present some new illustrative results.

In section \ref{sci} we briefly recall what Configuration Interaction (CI) methods are about and 
present the basic ideas of perturbatively selected CI approaches. We emphasize on the very high 
efficiency of SCI in approaching the exact Full CI limit using only a {\it tiny} fraction 
of the full Hilbert space of determinants.
Selecting important determinants being a natural idea,
it is no surprise that it has been introduced a long time ago and has been 
rediscovered many times under various forms since then.
To the best of our knowledge selected CI appeared
for the first time in 1969 in two independent works by Bender and Davidson\cite{bender_pr_1969}
and Whitten and Hackmeyer.\cite{Whitten_1969}
In practice, the flavor of SCI we employ is the CIPSI approach introduced 
by Malrieu and collaborators in 1973.\cite{huron_jcp_1973} CIPSI being our working algorithm 
for generating CI expansions, a brief description is given here.
It is noted that the recent FCI-QMC
method of Alavi {\it et al.}\cite{booth_jcp_2009, cleland_jcp_2010} is essentially a SCI 
approach, except that selection of determinants in FCI-QMC is done stochastically instead of
deterministically. 

In section \ref{results} the performance of CIPSI is illustrated for the case of the 
water molecule at equilibrium geometry using the cc-pCV$n$Z family of basis sets, with $n=2$ to 5 
and for the whole set of 55 molecules and 9 atoms of the G1 standard 
set.\cite{pople_jcp_1989,curtiss_jcp_1990} It is shown that in all cases the FCI limit 
is closely approached.

In section \ref{dmc} the use of CIPSI nodes in DMC is discussed. We first present our 
motivations and then comment on the key result observed, namely that in all applications 
realized so far the fixed-node error associated with the approximate nodes of the CIPSI 
expansion is found to 
systematically decrease both as a function of the number of selected determinants 
and as the size of the basis set. This remarkable property provides a convenient way of 
controlling the fixed-node error. Let us emphasize that in contrast with common practice in QMC
the molecular orbitals are not stochastically re-optimized here.
An illustrative application to the water molecule is presented.\cite{caffarel_jcp_2016}
Of course, the main price to pay is the need of using much larger CI 
expansions than usual. The main ideas of our recently proposed 
approach\cite{preprint_multidets} to handle very large number of determinants in QMC are presented.
In practice, converged DMC calculations using trial wavefunctions 
including up to a few millions of determinants are feasible. The computational increase with respect 
to single-determinant calculations is roughly proportional to $\sim \sqrt{N_{dets}}$ with a 
small prefactor.

In section \ref{pseudo} the implementation of effective core potentials (ECP) 
in DMC using CIPSI trial wavefunctions is presented. 
As already proposed some time ago,\cite{Hurley_1987,Hammond_1987}
CI expansions allow to calculate analytically the action of the nonlinear 
pseudo-potential operator on the trial wavefunction.
In this way, the use of quadrature points to integrate the wavefunction over the sphere as usually 
done\cite{mitas_jcp_1991}
is avoided and a gain in computational effort essentially proportional to the number of grid points is 
achieved. The effectiveness of the approach 
is illustrated in the case of the atomization energy of the C$_2$ molecule. 

Finally, Sec. \ref{conclu} presents a detailed summary of the main features 
of the DMC-CIPSI approach and some lines of research presently under investigation are mentioned.

\section{Selected Configuration Interaction}
\label{sci}

\subsection{Configuration Interaction methods}
In Configuration Interaction the wavefunction is written as a sum of Slater determinants
\begin{equation}
|\Psi\rangle = \sum_i c_i |D_i\rangle
\label{ciexp}
\end{equation}
where determinants are built over spin-orbitals. Let $\{ \phi_k \}$ be the set of $N_{\rm MO}$
orthonormal molecular orbitals used, the size of the full Hilbert space is given by the 
number of ways of distributing the $N_{\uparrow}$ electrons among the orbitals times the 
corresponding number for $N_{\downarrow}$ electrons.
The total size of the full CI space is then (no symmetries are considered)
$$
N_{FCI}= \binom{N_{\rm MO}}{N_{\uparrow}} \binom{N_{\rm MO}}{N_{\downarrow}}
$$
The CI eigenspectrum is obtained by diagonalizing the Hamiltonian matrix, 
$H_{ij}= {\langle D_i|H|D_j\rangle}$ within the orthonormal basis of determinants.
In practice, the exponential increase of the FCI space restricts the use of FCI 
to small systems including a small number of electrons and molecular orbitals ($N_{FCI}$
not greater than about $10^9$). To go beyond, the FCI expansion has to be truncated. 
The most popular strategy consists in defining a subspace of determinants chosen {\it a priori}.
Typically, the Hartree-Fock determinant (or a few determinants)
is chosen as reference and all possible determinants built by promoting a given number of electrons 
from the HF occupied orbitals to the virtual ones are considered.
In the CIS approach only single excitations are considered, 
in CISD all single and double excitations, etc.

\begin{figure}[h!]
\begin{center}
\includegraphics[width=1.\columnwidth]{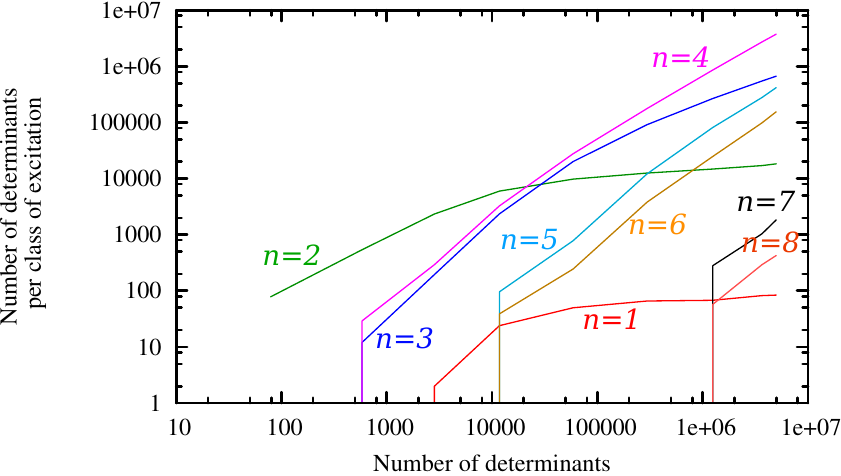}
\caption{N$_2$ in the cc-pVTZ basis set (R$_{N-N}$=1.0977 \AA). Variation of the number 
of determinants with $n$-excitations with respect to the Hartree-Fock determinant 
in the CIPSI expansion as a function of the number of selected determinants up to $5 \times 10^6$.
}
\label{n2}
\end{center}
\end{figure}

Now, numerical experience shows that among all possible determinants corresponding to 
a given number of excitations, only a {\it tiny} fraction 
plays a significant role in constructing the properties of the low-lying eigenstates. Furthermore, 
the weight of a determinant in the CI expansion is not directly related 
to its degree of excitation. For example, quadruply-excited 
determinants may play a more important role than some doubly- or singly-excited determinants. However, 
in practice, limiting the maximum number of excitations to about six is usually sufficient to 
get chemical accuracy. To give some quantitative illustration of these statements, Figure \ref{n2}
presents the number of determinants per class of excitations $n$ as a function of the number of 
determinants in the CIPSI wavefunction for the $N_2$ molecule 
at equilibrium geometry (cc-pVTZ basis set). Without entering now into the details 
of CIPSI presented below, let us just note that for 5$ \times 10^6$ determinants the CIPSI expansion 
has almost converged to the FCI solution. Accordingly, results presented in the figure 
for the distribution of excitations is essentially that of the FCI wavefunction.

As a consequence of the preceding remarks, it is clear that it is desirable to find a way 
of selecting only the most important determinants of the FCI expansion without considering all 
those of negligible weight (the vast majority). This is the purpose 
of selected configuration interaction approaches. 

\subsection{Selected CI and CIPSI algorithm}
To the best of our knowledge Bender and Davidson\cite{bender_pr_1969} and 
Whitten and Hackmeyer\cite{Whitten_1969} were the first in 1969 to 
introduce and exploit the idea of selecting determinants in CI approaches.
In their work Bender and Davison proposed to select space configuration using an energy 
contribution criterion.
Denoting $|\phi_0\rangle$ the restricted HF CSF-configuration,
$|\phi_i^l \rangle$ all possible spin configurations issued from the 
space configuration,
and 
\begin{equation}
\epsilon_i^{(2)} = \frac{1}{k} \sum_{l=1}^k \frac{ {|\langle \phi_i^l |H|\phi_0 \rangle |}^2 }
{ \langle \phi_0 |H|\phi_0 \rangle - \langle \phi_i^l |H|\phi_i^l \rangle }
\end{equation}
the ``average'' perturbative energy contribution, the space configurations were ordered according 
to this contribution and those determinants contributing the most selected. 
The CI wavefunction was then constructed 
by using the selected configurations, $|\phi_0\rangle$, and 
all single excitations. A few months later, a similar idea using the very same perturbative criterion 
was introduced independently by Whitten and Hackmeyer.\cite{Whitten_1969} In addition, they 
proposed to improve step-by-step the CI expansion by iterating the selection step to reach 
the most important determinants beyond 
double-excitations.

In 1973 Malrieu and collaborators\cite{huron_jcp_1973} presented the CIPSI method (and later on 
an improved version of it\cite{cipsi_1983}). In CIPSI the construction of the multirefence 
variational space is essentially identical to that of Whiten and Hackmeyer. 
However, in order to better describe the dynamical correlation effects poorly reproduced by 
the multireference space, a perturbational calculation of the remaining 
correlation contributions was proposed. In applications the perturbational part is usually important 
from both a qualitative and quantitative point of view.\\

The CIPSI algorithm being our practical scheme for generating selected CI expansions, 
let us now present its main steps.\\

$\bullet$ Step 0: Start from a given determinant ({\it e.g.} the Hartree-Fock determinant) or set of determinants, thus defining
an initial reference subspace: $S_0=\{|D_0\rangle,...\}$. Diagonalize $H$ within $S_0$ and get the ground-state energy $ E_0^{(0)}$ and eigenvector:
\begin{equation}
|\Psi_0^{(0)}\rangle= \sum_{i \in S_0} c_i^{(0)} |D_i \rangle 
\end{equation}
Here and in what follows, a superscript on various quantities is used to 
indicate the iteration number.\\

Then, do iteratively ($n=0,...$):\\

{$\bullet$ Step 1}: Collect all {\it different} determinants $|D_{k}\rangle$ connected by 
$H$ to $|\Psi_0^{(n)}\rangle$, that is 
\begin{equation}
\langle \Psi_0^{(n)}|H|D_{k}\rangle \ne 0
\end{equation}
and not belonging to the reference space $S_{n}$.\\

{$\bullet$ Step 2}: Compute the small energy change of the total energy due to each connected 
determinant as evaluated at second-order perturbation theory 
\begin{equation}
\delta e(|D_{k}\rangle)=-\frac{{|\langle \Psi_0^{(n)}|H|D_{k}\rangle|}^2}{H_{kk}-E_0^{(n)}}
\label{e2pert}
\end{equation}

{$\bullet$ Step 3}: Add the determinant $|D_{k^*}\rangle$ associated with the largest $|\delta e|$ to the reference subspace: 
$$S_{n} \rightarrow S_{n+1}= S_{n} \cup \{|D_{k*} \rangle\}$$\\
Of course, instead of adding only one determinant a group of determinants can be selected using 
a threshold. This is what is actually done in practice.\\

{$\bullet$ Step 4}: Diagonalize $H$ within $S_{n+1}$ to get: 
\begin{equation}
|\Psi_0^{(n+1)}\rangle= \sum_{i \in S_{n+1}} c_i^{(n+1)} |D_i\rangle \;\;\; {\rm with}\;\;\; E_0^{(n+1)}
\label{psi_cip}
\end{equation}

{$\bullet$} Go to step 1 or stop if the target size for the reference subspace has been reached.\\

Denoting $N_{\rm dets}$ the final number of determinants, the resulting ground-state 
$|\Psi_0(N_{\rm dets})\rangle$ is the variational CIPSI solution. It is the expansion
used in DMC to contruct the determinantal part of the trial wavefunction.

A second step in CIPSI is the calculation of a perturbational estimate of the correlation energy left
between the variational CIPSI energy and the exact FCI one. At second order, this 
contribution writes
\begin{equation}
E_{PT2}= - \sum_{k \in \mathcal{M}} \frac{  { {|\langle \Psi_0(N_{\rm dets})|H|D_k} \rangle|}^2}{H_{kk}- E_0(N_{\rm dets})}
\label{pt2}
\end{equation}
where $\mathcal{M}$ denotes the set of all determinants not belonging to the reference space 
and connected to the CIPSI expansion $|\Psi_0(N_{\rm dets})\rangle$
by $H$ (single and double excitations only) and $E_0(N_{\rm dets})$ the variational CIPSI 
energy. In practice, this contribution allows to recover a major part of the remaining correlation 
energy.\\

At this point a number of remarks are in order:\\

i.) Although the selection scheme is presented here for computing the ground-state eigenvector only, no special difficulties
arise when generalizing the scheme to a finite number of states (see, {\it e.g.}\cite{cipsi_1983})\\

ii.) The decomposition of the Hamiltonian $H$ underlying 
the perturbative second-order expression introduced in step 2 is 
known as the Epstein-Nesbet 
partition.\cite{en1,en2} This decomposition is not unique, other possible choices are the M{\o}ller-Plesset partition\cite{mp} or the barycentric one,\cite{huron_jcp_1973} 
see discussion in \cite{cipsi_1983}.\\

iii.) Instead of calculating the energetic change perturbatively, expression (\ref{e2pert}), it can be preferable to employ
the non-perturbative expression resulting from the diagonalization of $H$ into the two-dimensional basis consisting of 
the vectors $|\Psi_0^{(n)}\rangle$ and $|D_{k}\rangle$. Simple algebra shows that the energetic change is given by
\begin{equation}
\delta e(|D_k\rangle)= 
      \frac{1}{2} \left[H_{kk} - E_0(N_{\rm dets})\right]
      \left[1-\sqrt{1 + \frac{4 {|\langle \Psi_0^{(n)}|H|D_k \rangle|}^2}{{[H_{kk}-E_0(N_{\rm dets})]}^2}}\right]
\label{e2ex}
\end{equation}
In the limit of small transition matrix elements, $\langle \Psi_0^{(n)}|H|D_k \rangle$, both expressions (\ref{e2pert}) and (\ref{e2ex}) coincide.
The non-perturbative formula is used in our applications.\\

iv.) The implementation of this algorithm can be performed using limited amount of central memory. On the other hand, the CPU time required 
is essentially proportional to $N_{\rm dets} N_{\rm occ}^2 N_{\rm virt}^2$ where $N_{\rm occ}$ is the number of occupied molecular 
orbitals and $N_{\rm virt}$ the number of virtual orbitals.

\subsection{Selected CI variants}
As already pointed out selecting the most important determinants of the FCI 
expansion is a so natural idea that, since the pioneering work of Bender and 
Davidson\cite{bender_pr_1969} and Whitten 
and Hackmeyer,\cite{Whitten_1969} several variants of SCI approaches have been proposed. 
In practice, the actual differences 
between approaches are usually rather minor and most ideas and technical aspects seem to have 
been re-discovered several times by independent groups. To give a fair account of the subject 
and an exhaustive list of references is thus difficult. Here, we limit ourselves to 
the references we are aware of, namely
\cite{bender_pr_1969,Whitten_1969,Hackmeyer_1971,langhoff_ijqc_1973,huron_jcp_1973,
buenker_1974,buenker_1975,buenker_1978,bruna_1980,buenker_1981,cipsi_1983,cimiraglia_jcp_1985,
cimiraglia_jcc_1987,harrison_1991,cimiraglia_ijqc_1996,Angeli_1997_I,Angeli_1997_II,Angeli_2001,Bunge_2006,Roth_2007,Roth_2009,mcci,evangelista_jcp_2014,tubman_2016,tubman_cyrus_2016}.
Regarding more specifically CIPSI, there has been a sustained research activity 
conducted during the 80's and 90's by research groups in Toulouse (Malrieu and coll.), 
Pisa (Angeli, Persico, Cimiraglia and coll.), and then Ferrara (Angeli, Cimiraglia) including the 
development at Pisa of a very efficient CIPSI code 
using diagrammatic techniques\cite{cipsi_code,cimiraglia_jcp_1985,cimiraglia_ijqc_1996}. 
Thanks to all this, 
CIPSI has been extensively applied for years by several groups to a variety of 
accurate studies of ground and excited states and potential energy
surfaces (see, for example 
\cite{Povill_1992,Illas_1991a,Illas_1991,Millie_1986,Persico_1991,Illas_1988,Cabrol_1996,Cabrol_1996,
Angeli_1996,Milli__2000,M_dl_1997,Cattaneo_1999,Li_2011,Mennucci_2001,Novoa_1988,Aymar_2006,Aymar_2005})
Finally, note that in the last years our group has developped its own CIPSI code,
Quantum Package. This code has been designed to be 
particularly easy to install, run and modify; it can be freely downloaded at \cite{quantum_package}.

\subsection{FCI-QMC as a stochastic selected CI approach}
Full Configuration Interaction Quantum Monte Carlo (FCI-QMC) is a method for solving stochastically 
the FCI equations.\cite{booth_jcp_2009, cleland_jcp_2010} Introducing as in DMC an imaginary time $t$ the coefficients $c_i$ of 
the CI expansion, Eq.(\ref{ciexp}), are evolved in time using the operator $[1-\tau (H-E)]$ as small-time propagator
\begin{equation}
{\bf c}(t+\tau) =[1-\tau(H-E)] {\bf c}(t)
\label{ele}
\end{equation}
${\bf c}$ being the vector of coefficients, $E$ some reference energy, and $\tau$ the time step.
After $n$ steps the coefficients are given by
\begin{equation}
{\bf c}(t) = [1-\tau(H-E)]^n {\bf c}(t=0).
\label{eqFCI}
\end{equation}
In the long-time limit ($t=n\tau$ large) the vector ${\bf c}$ converges 
to the exact CI vector (independently on initial conditions ${\bf c}$(t=0) provided that 
$\langle {\bf c}$(t=0)$|{\bf c}\rangle \ne 0$
and for a sufficiently small time step).
As in all QMC methods, a set of walkers is introduced for sampling coefficients and 
a few simple stochastic rules realizing {\it in average} the action of $H$ according to 
Eq.(\ref{ele}) are introduced (spawning, death/cloning and annihilation).
Note that equations of evolution (\ref{eqFCI}) are similar to those of continuous DMC 
(electrons moving in ordinary space) where a small-time expression of operator 
$e^{-\tau (H-E)}$ is used, and are essentially identical to the equations of lattice DMC 
(see {\it e.g.},\cite{van_Bemmel_1994})
The two main differences of FCIQMC with other QMC approaches are 
the fact that no trial vector is introduced (thus, avoiding the fixed-node error) 
and that the stochastic rules used are particularly efficient in attenuating 
the sign instability inherent to all stochastic simulations of fermionic systems
(annihilation at each MC step of walkers of opposite sign on occupied determinants
and use of the initiator approximation).

At a given time $t$ the CI expansion is stochastically realized by the distribution 
of walkers as 
$$
|\Psi\rangle = \sum_i  n_i |D_i\rangle
$$
where $n_i$ is the sum of the signed weight of walkers on Slater determinant $|D_i\rangle$ 
($M=\sum_i |n_i|$= total number of walkers). 
This wavefunction is the counterpart of the CIPSI expansion at iteration $n$, Eq.(\ref{psi_cip}).
As in CIPSI at the next step $t+\tau$ (next iteration $n+1$) new determinants will appear. 
In FCI-QMC it is realized through spawning.
Some determinants may also disappear through the action 
of the diagonal part of the Hamiltonian $[1-\tau (H_{ii}-E)]$ (death/cloning step).  These two 
steps are designed to reproduce in average the action of the propagator on determinant $D_i$
$$
[1-\tau (H-E)]|D_i\rangle = [1-\tau (H_{ii}-E)]|D_i\rangle -\tau  \sum_{k \ne i}
H_{ik}|D_k\rangle.
$$
In CIPSI a given determinant $|D_i\rangle$ is selected only once during 
iterations via Eq.(\ref{e2pert}). 
In latter iterations it is included in the reference space and does not participate 
anymore to the selection. Starting from some initial determinant (usually the HF determinant) 
the probability of selecting $|D_i\rangle$ at some given iteration $n$ is related to the 
existence of a series of $(n-1)$ intermediate determinants $(|D_{i_1}\rangle,|D_{i_2}\rangle,
...,|D_{i_k}\rangle,...)$ different from $|D_i\rangle$
and connecting it to the initial determinant so that the product 
$$
\prod_k \frac{|H_{i_{k+1}i_k}|^2}{H_{i_{k+1}i_k}-E_0}
$$
is large compared to products corresponding to other series of intermediate determinants.

In FCI-QMC a determinant $|D_i\rangle$ is spawned (selected) from $|D_j\rangle$ 
according to the magnitude of $H_{ii}$ and -in contrast with CIPSI- with no direct 
dependence on the inverse of $(H_{ii}-E_0)$.
However, during MC iterations the number of walkers on a given determinant evolves in time    
according to the death/cloning step and leads to a weighted contribution of 
determinants to spawning. After integration in time the weight of the determinants $|D_i\rangle$
can be estimated to be
about $\int dt e^{-t(H_{ii}-E_0)}$ that is, 
$\sim \frac{1}{H_{ii}-E_0}$ for large enough time.
As seen, FCI-QMC and CIPSI are in close connection.

\section{Applications of CIPSI}
\label{results}
\subsection{The water molecule}
\label{cipsi_water}
\begin{figure}[h!]
\begin{center}
\includegraphics[angle=-90,width=0.9\columnwidth]{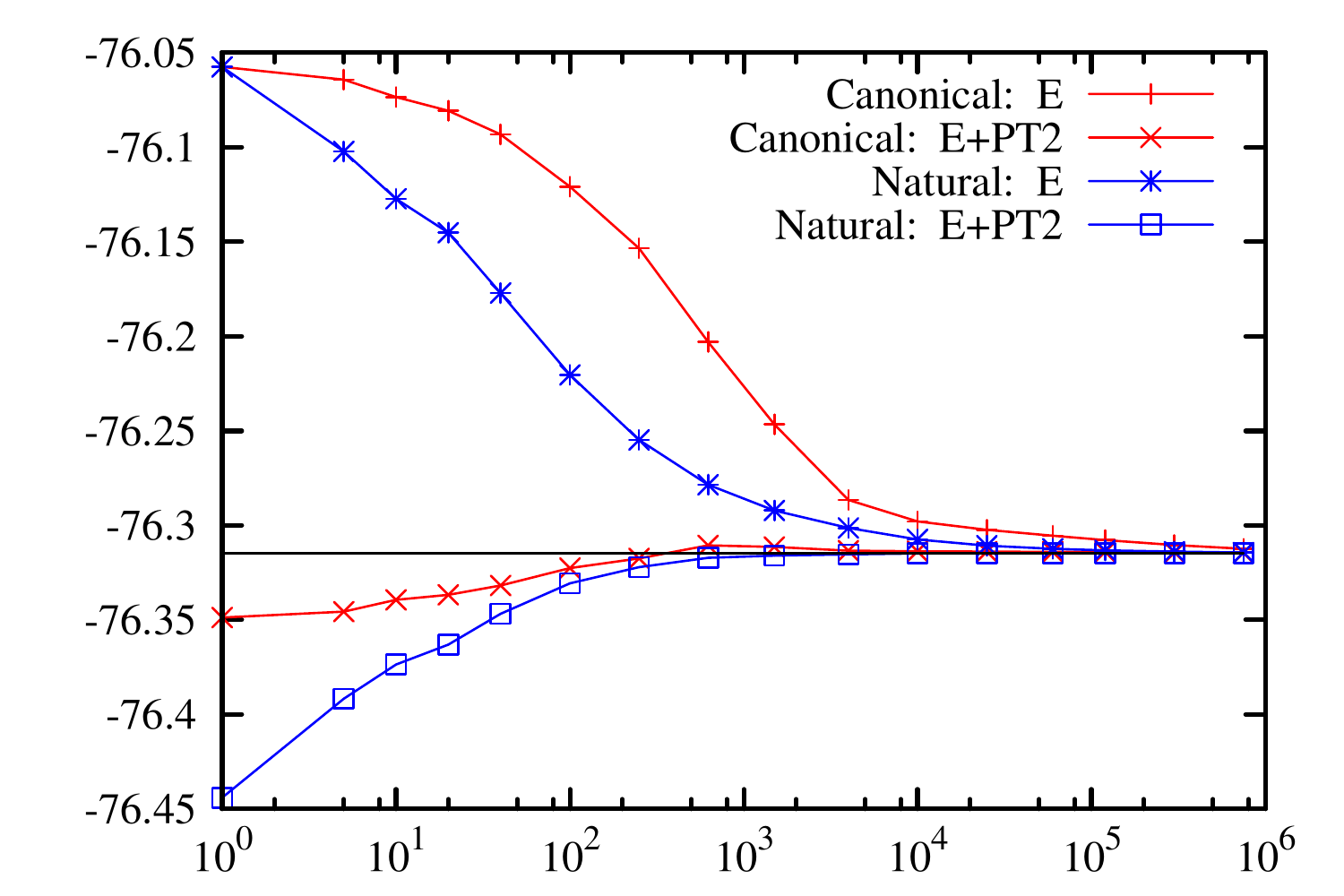}
\caption{Energy convergence of the variational and full CIPSI energies as a function 
of the number of selected determinants using canonical and natural orbitals. Energy in a.u.}
\label{chan}
\end{center}
\end{figure}

To exemplify CIPSI all-electron calculations for the water molecule 
using basis sets of various sizes are presented. In our first example we propose to reproduce 
the density matrix renormalization group (DMRG) calculation of Chan and Head-Gordon\cite{Chan_2003} 
at geometry ($R_{OH}=1 \AA,\theta_{OH}=104.5^{\circ}$) and using the 
``Roos Augmented Double Zeta ANO'' 
basis set consisting of 41 orbitals\cite{Schuchardt_2007,JCC:JCC9}. The 
full CI Hilbert space contains about $5.6\;10^{11}$ determinants (no spin or space symmetries 
taken into account).
Calculations have been carried out using our perturbatively selected CI program
Quantum Package.\cite{quantum_package}
The energy convergence as a function of the number of selected determinants in
different situations is presented in Figure \ref{chan}.

Four different curves are shown together with the DMRG energy value of -76.31471(1)
of Chan and Head-Gordon\cite{Chan_2003} (solid horizontal line).
The two upper curves represent the CIPSI variational energy 
as a function of the number of selected determinants up to 
750 000 using either canonical or natural molecular orbitals. 
Natural orbitals have been obtained by
diagonalizing the first-order density matrix built with the largest expansion 
obtained using canonical orbitals.
As seen the convergence of both variational energies is very rapid. Using canonical orbitals 
an energy of -76.31239 a.u. is obtained with 750 000 determinants, a value differing from the 
FCI one by only 2.3 millihartree (about 1.4 kcal/mol).
As known the accuracy of CI calculations is significantly enhanced 
when using natural orbitals.\cite{Davidson1972235} Here, it is clearly the case and 
the lowest energy reached is now -76.31436 a.u. with an error of 0.35 millihartree (about 
0.2 kcal/mol). When adding the second-order energy correction $E_{PT2}$, Eq.(\ref{pt2}), 
the energy convergence is much improved (two lower curves of Figure \ref{chan}).
The kcal/mol (chemical) accuracy is reached 
with only 1000 and 4000 determinants using canonical and natural orbitals, respectively.
The best CIPSI energy including second-order correction and obtained with canonical orbitals 
is -76.31452 a.u. When using natural orbitals
the energy is found to converge with five decimal places to the value of -76.31471 a.u.,
in perfect agreement with the DMRG result of Chan and Head-Gordon, -76.31471(1) a.u.

Let us emphasize that approaching the FCI limit with such a level of accuracy and so few 
determinants (compared to the 
total number of $5.6\;10^{11}$) is particularly striking and is one of the most remarkable 
features of SCI approaches.

To illustrate the possibility of making calculations with much larger basis sets,
results obtained with 
the correlation-consistent polarized core-valence basis sets, cc-pCV$n$Z,
with $n$ going from 2 to 5 are presented. The geometry chosen is now the experimental 
equilibrium geometry, $R_{OH}=0.9572$ \AA {} and $\theta_{OH}=104.52^{\circ}$. 
The number of basis set functions are 28, 71, 174
and 255 for cc-pCVDZ, cc-pCVTZ,cc-pCVQZ, and cc-pCV5Z, respectively. 
The total number of determinants of the FCI 
Hilbert space with such basis sets are 
about $1\,10^{10}$,$1.7\,10^{14}$,$1.6\,10^{18}$, and $7.5\,10^{19}$, 
respectively.
\begin{figure}[!h]
\begin{center}
\includegraphics[width=1.\columnwidth]{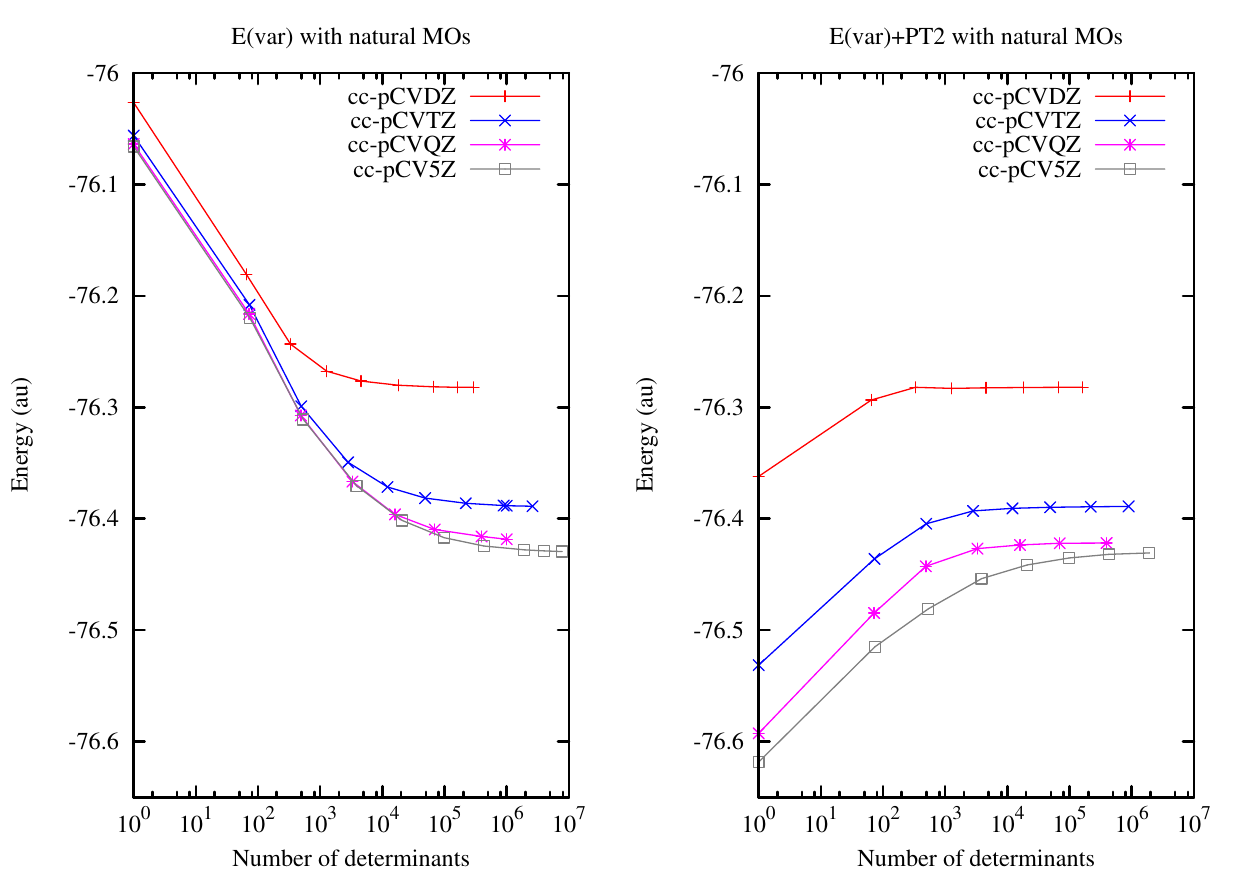}
\caption{Convergence of the energy with the number of selected determinants
(logarithmic scale).  The graph on the left displays the variational energy,
and the graph on the right shows the energy with the perturbative correction, Eq.(\ref{pt2}).}
\label{ecipsi}
\end{center}
\end{figure}
On the left part of Figure \ref{ecipsi} the convergence of the ground-state variational energy 
obtained for each basis set is shown. 
As seen, the convergence is still possible with such larger basis sets.
On the right part, 
the full CIPSI energy curves $(E_{var} + E_{PT2})$ are presented; each curve is found to converge with a good accuracy 
to the full CI limit.

\subsection{Generalization: The G1 set}

\begin{figure}[!h]
\begin{center}
\includegraphics[width=1.1\columnwidth]{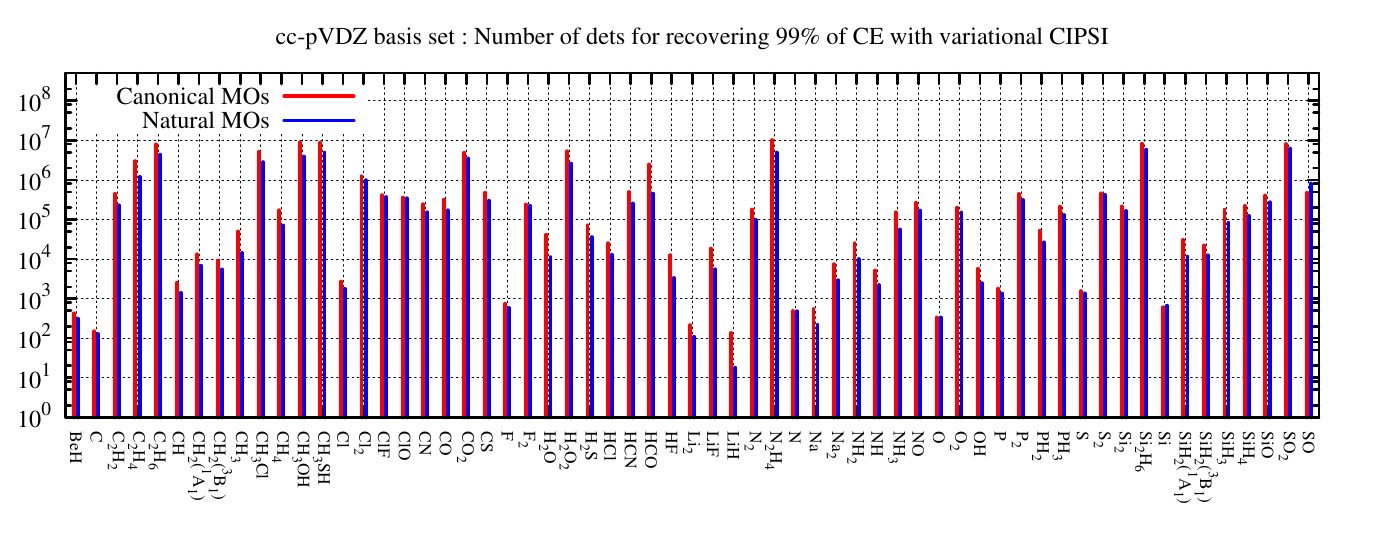}
\caption{Number of selected determinants required to recover 99\% 
of the total correlation energy at CIPSI/cc-pVDZ variational level. Results for canonical and natural orbitals
are given.}
\label{g1_1}
\end{center}
\end{figure}

\begin{figure}[!h]
\begin{center}
\includegraphics[width=1.1\columnwidth]{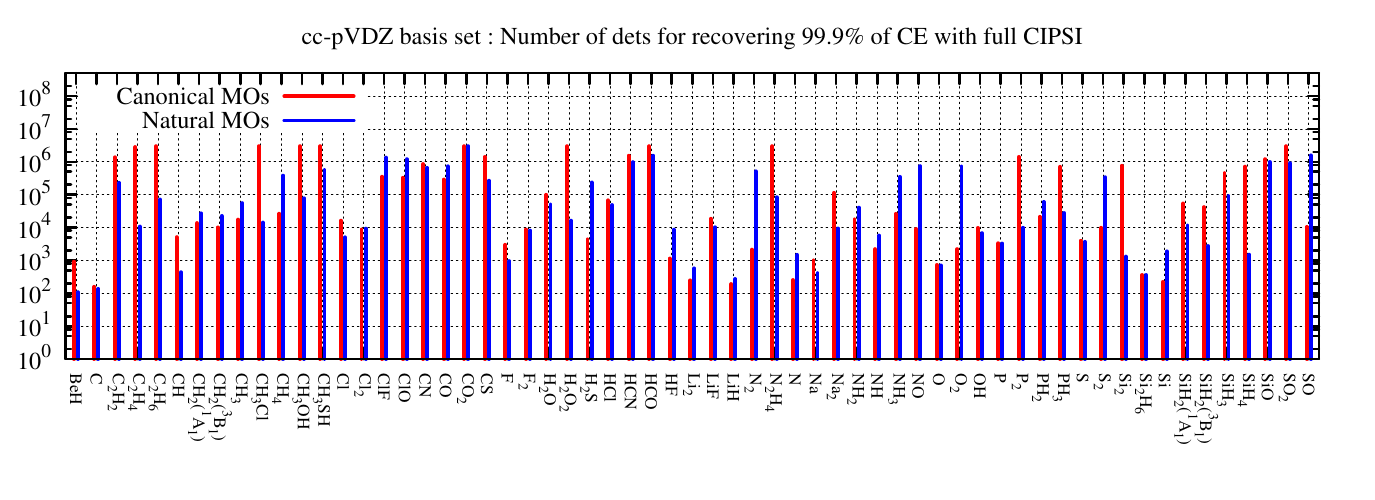}
\caption{Number of selected determinants required to recover 99.9\% 
of the total correlation energy at full CIPSI/cc-pVDZ level $(E_{var} + E_{PT2})$. 
Results for canonical and natural orbitals
are given.}
\label{g1_2}
\end{center}
\end{figure}

\begin{figure}[h!]
\begin{center}
\includegraphics[width=1.\columnwidth]{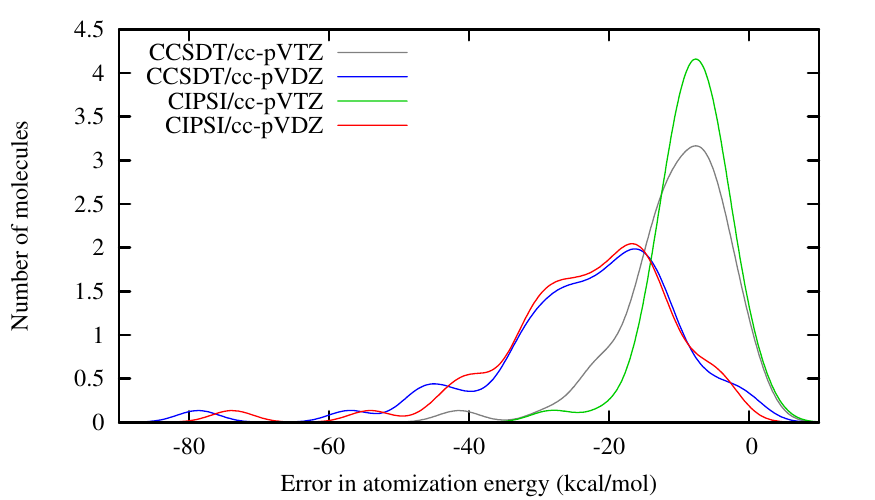}
\caption{Distribution of errors in atomization energies for the whole G1 set of atomic and molecular 
systems calculated with CIPSI and CCSD(T). Results shown for cc-pVDZ and cc-pVTZ basis sets.}
\label{ccsdt}
\end{center}
\end{figure}

In contrast with the exact Full-CI approach which takes into account 
the entire set of determinants and is thus rapidly unfeasible, 
CIPSI can be used for much larger systems. The exact limits depend of course on the size 
of the basis set used, the number of electrons, and also on the level of convergence asked for when 
approaching the full CI limit. To illustrate the feasibility of CIPSI for larger systems 
we present systematic all-electron calculations for the G1 benchmark set of Pople 
and collaborators.\cite{g2} The set is composed of 55 molecules and 9 different atoms. 
The cc-pVDZ and cc-pVTZ basis sets have been used. For all systems and both basis sets a quasi-FCI 
convergence has been reached. 
In Figure \ref{g1_1} the number of selected determinants needed 
to recover 99\% of the correlation energy at CIPSI variational level (cc-pVDZ basis set) is 
plotted for each molecule or atom. For each system results are given either for 
canonical or natural orbitals.
Depending on the importance of the multiconfigurational character of the system, 
this number may vary considerably (from a few tens to about 10$^7$).
As expected, the number of determinants needed using natural orbitals is most of the times smaller 
and sometimes comparable. Figure \ref{g1_2} is similar to the preceding figure, except that numbers 
are given now for a full CIPSI calculation including the second-order energy correction and that 
a much greater accuracy corresponding to 99.9\% of the correlation energy is targeted.
As seen, it is remarkable that such a high precision can be reached for all systems with a number of 
determinants not exceeding $\sim 10^7$. In contrast with variational 
calculations, it should be noted that the use of natural orbitals does not systematically 
improve the convergence.
Finally, some comparison with accurate CCSD(T) 
calculations performed using the same basis sets and geometries are presented. In Figure \ref{ccsdt} 
the distribution of errors in atomization energies calculated with both CCSD(T) and CIPSI methods are 
plotted. For the cc-pVDZ basis set, CCSD(T) and CIPSI curves are very similar, indicating that 
CCSD(T) calculations have also reached the quasi full CI limit. For the larger cc-pVTZ basis set, 
the two curves remain similar but some significant differences show up with 
CIPSI results more distributed toward small errors due to a better description of multireference 
systems.

\section{Using CIPSI nodes in DMC}
\label{dmc}
\subsection{Motivations}
In DMC the standard practice is to introduce compact trial wavefunctions 
reproducing as much as possible the mathematical and physical properties of 
the exact wave function. Next, the ``best'' nodes are determined through optimization 
of the parameters of the trial wavefunction in a preliminary variational Monte Carlo (VMC) run. 
The objective function to minimize is either 
the variational energy associated with the trial wavefunction or the variance of 
the Hamiltonian (or a combination of both). A number of algorithms have been
elaborated to perform this important practical step as 
efficiently as possible.\cite{Filippi_2000,Schautz_2002,Umrigar_2005,Scemama_2006,Toulouse_2007,Toulouse_2008} 
No limitations existing in QMC for the choice of 
the functional form of the trial wavefunction, many different expressions have been 
introduced (see, {\it e.g.} \cite{mosko,sorella,mitas,rios,goddard,fili_vb,braida,bouabca}).
However, the most popular one is certainly the Jastrow-Slater trial wavefunction  
expressed as a short expansion over a set of Slater determinants
multiplied by a global Jastrow factor describing explicitly the
electron-electron and electron-electron-nucleus interactions and,
in particular, imposing the electron-electron cusp conditions 
associated with the zero-interelectronic distance limit of the true wavefunction.

In the DMC-CIPSI approach the determinantal part of the trial wavefunction 
is built using systematic CIPSI expansions.
The main motivation is that CI approaches provide a simple, deterministic, and systematic way
of constructing wavefunctions of controllable quality. In a given one-particle basis set,
the wavefunction is improved by increasing the number of determinants, up to the Full
CI (FCI) limit. Then, by increasing the basis set, the wavefunction can be further improved,
up to the complete basis set (CBS) limit where the exact solution of the continuous 
electronic Schr{\"o}dinger equation is reached. The CI nodes, which are defined as the zeroes of 
the expansion, are also expected to follow such a systematic improvement, thus 
facilitating the control of the fixed-node error. A second important motivation is that 
the stochastic optimization step can be avoided since a systematic way of improving 
the wavefunction is now at our disposal. The optimal CI coefficients 
are obtained by the (deterministic) diagonalization of the Hamiltonian matrix 
in the basis set of Slater determinants. It is a simple and robust step which 
leads to a unique set of coefficients. Furthermore, it can be made automatic,
an important feature in the perspective of designing a fully black-box QMC code.
Finally, using {\it deterministically} constructed nodal structures greatly facilitates 
the use of nodes evolving {\it smoothly} as a function of any parameter of the Hamiltonian. 
It is important when calculating potential energy surfaces (see, our application to the 
F$_2$ molecule,\cite{giner_jcp_2015}) or 
response properties under external fields.

The main price to pay for such advantages is of course the need of
considering much larger multideterminant expansions (from tens of thousands up to a few millions)
than in standard DMC implementations where compactness of the trial wavefunction is searched for.
However, efficient algorithms
have been proposed to perform such calculations\cite{Nukala_2009,Clark_2011,Weerasinghe_2014}. 
Very recently, we have also presented an efficient algorithm 
for computing very large CI expansions. Its main ideas
are briefly summarized in section \ref{largedets} below.

\subsection{Toward a better control of the fixed-node approximation}

\begin{figure}[h!]
\includegraphics[width=0.9\textwidth]{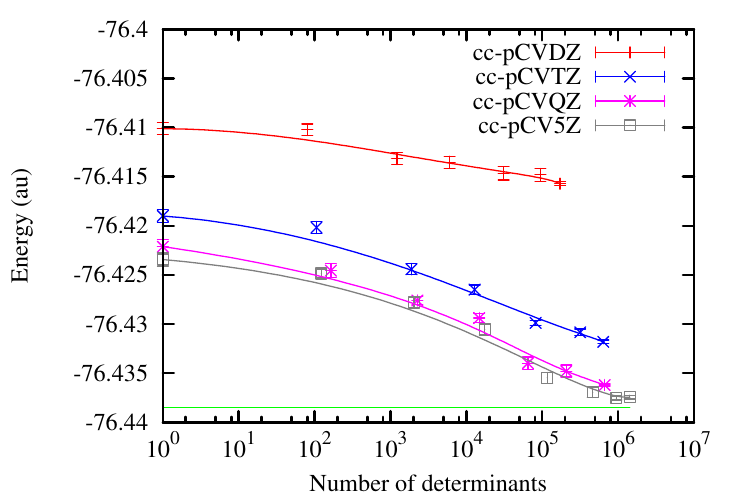}
\caption{DMC energy of the water molecule as a function of the number of determinants in the trial wave function (logarithmic scale). The horizontal solid line indicates the experimentally derived 
estimate of 
the exact nonrelativistic energy.\cite{klopper_mp_2001}}
\label{fig:h2o_dmc}
\end{figure}

A remarkable property systematically observed so far in our DMC applications using
large CIPSI expansions\cite{giner_2013,scemama_jcp_2014,giner_jcp_2015} is that,
except for a possible transient regime at small number of determinants,\cite{note2} the fixed-node
error resulting from the use of CIPSI nodes is found to decrease monotonically, 
both as a function of the number of selected determinants, $N_{dets}$, and of the basis set size, $M$. 
This result is illustrated here in the case of the water 
molecule at equilibrium geometry. Results shown here complement our recent 
benchmark study on water.\cite{caffarel_jcp_2016}. 
In Figure \ref{fig:h2o_dmc} all-electron fixed-node
energies obtained with DMC-CIPSI as a function of the number of selected determinants for the first 
four cc-pCVnZ basis set (n=2-5) are reported. Calculations have been performed using the 
variational CIPSI expansions of the preceding 
subsection. In practice, DMC simulations have been realized using our general-purpose
QMC program QMC=Chem (downloadable at \cite{qmcchem}). A minimal Jastrow 
prefactor taking care of the electron-electron cusp condition is employed and molecular orbitals
are slightly modified at very short electron-nucleus distances to impose exact 
electron-nucleus cusp conditions. The time step used, $\tau= 2 \times 10^{-4}$ a.u., 
has been chosen small enough to make the finite time step error not observable 
with statistical fluctuations. As seen on the figure the convergence of DMC energies both 
as a function of the number of determinants and of the basis set are almost reached. 
The value of $-76.43744(18)$ a.u. obtained with the largest basis set and 1 423 377 determinants 
is, to the best of our knowledge, the lowest upper bound reported so far, 
the experimentally derived estimate of the exact nonrelativistic energy 
being -76.4389(1) a.u.\cite{klopper_mp_2001}
Thanks to our recent algorithm for calculating very large number 
of determinants in DMC\cite{preprint_multidets} (see, section \ref{largedets} below), 
the increase of CPU time 
for the largest calculation including more than 1.4 million of determinants 
compared to the same calculation limited to the Hartree-Fock determinant
is only $\sim 235$.

In practice, the possibility of calculating fixed-node energies displaying such a regular behavior 
as a function of the number of determinants and molecular orbitals 
is clearly attractive in terms of 
control of the fixed-node error. For example, in our benchmark study of the water
molecule\cite{caffarel_jcp_2016} 
it was possible to extrapolate the DMC energies obtained with each cc-pCVnZ basis set
as a function of the cardinal number $n$, as routinely 
done in deterministic CI calculations. Using a standard $1/n^3$ law a very accurate 
DMC-CIPSI energy value of -76.43894(12) a.u. was obtained, in full agreement with the estimate 
exact value of -76.4389(1) a.u.\cite{caffarel_jcp_2016}

At this point, we emphasize that the observed property of systematic decrease of the energy 
as a function of the number of determinants is known
not to be systematically true for a general CI expansion (see,
{\it e.g.} \cite{flad_book_1997}). Here, its validity may probably be attributed to the fact that
determinants are selected in a hierarchical way (the most important ones first),
so that the wavefunctions quality increases step by step, and so the quality of nodes.
However, from a mathematical point of view, such a property 
is far from being trivial. There is no simple argument why the FCI nodes obtained from 
minimization of the {\it variational} energy with respect to the multideterminant coefficients 
would lead to the best nodal structure
(minimum of the {\it fixed-node} energy with respect to such coefficients). In a general 
space (not necessarily a Hilbert space of determinants) 
it is easy to construct a wavefunction of poor quality having a high variational 
energy but exact nodes and, then, to exhibit a wavefunction with a much lower energy but wrong nodes. 
To demonstrate the validity or not of the observed property in a finite space of determinants 
built with molecular orbitals expanded in a finite basis set remains to be done.

\subsection{Evaluating very large number of determinants in QMC}
\label{largedets}
The algorithm we use to run DMC calculations with a very large number of determinants 
(presently up to a few millions) has been presented in detail in \cite{preprint_multidets}. 
Its efficiency is sufficiently high to 
perform converged DMC calculations with a number of determinants up to a few millions 
of determinants. In the case of the chlorine atom discussed in \cite{preprint_multidets} 
a trial wavefunction including about 750 000 determinants has been used with a computational 
increase of about 400 compared to a single-determinant calculation. As already mentioned above,
in the benchmark calculation of the water molecule\cite{caffarel_jcp_2016} 
up to 1 423 377 determinants have been used for a computational increase of only $\sim$ 235.\\

The main ideas of the algorithm are as follows.\\

$\bullet$ {\it O($\sqrt{N_{dets}}$)-scaling}. A first observation is that the determinantal 
part of trial wavefunctions built 
with $N_{dets}$ determinants can be rewritten as a function of a set of {\it different} 
{\it spin}-specific 
determinants $D^\sigma_i({\bf R}_\sigma)$ ($\sigma=\uparrow,\downarrow$)
as follows
\begin{equation} \label{eq:main}
\Psi_{Det}({\bf R})  = \sum_{i=1}^{N_{\rm dets}^\uparrow} \sum_{j=1}^{N_{\rm dets}^\downarrow} C_{ij} 
D^\uparrow_i({\bf R}_\uparrow) D^\downarrow_j({\bf R}_\downarrow)  
\end{equation}
where ${\bf C}$ is a matrix of
coefficients of size $N_{\rm dets}^\uparrow \times N_{\rm dets}^\downarrow$,
${\bf R}=({\bf r}_1,...,{\bf r}_N)$ denotes the full set of
electron space coordinates, and ${\bf R}_\uparrow$ and ${\bf R}_\downarrow$ the
two subsets of coordinates associated with $\uparrow$ and $\downarrow$ electrons.

In standard CI expansions the number of unique 
spin-specific determinants is much smaller than $N_{dets}$ and typically scales as $\sqrt{N_{dets}}$.
It is true for FCI expansions where all possible determinants are considered.
Indeed, $N_{\rm dets}^\sigma$ attains its maximal value of $\binom {N_{\rm MO}}{N_\sigma}$
and since $N_{\rm dets}$ is given as $N_{\rm dets}^\uparrow \times N_{\rm dets}^\downarrow$, 
the number of unique spin-specific determinants $D^\sigma({\bf R})$ 
is of order $\sqrt{N_{\rm dets}}$. However, it is in general also true 
for the usual truncated expansions (CASSCF, CISD, etc.) essentially because the 
numerous excitations implying multiple excitations of spin-like electrons plays a marginal role 
and have a vanishing weight.\\

$\bullet$ {\it Optimized Sherman-Morrison updates}. As proposed in a number of works,\cite{Nukala_2009,Clark_2011,Weerasinghe_2014}
we calculate the determinants and their derivatives
using the Sherman-Morrison (SM) formula for updating the inverse Slater matrices.
However, in contrast with other implementations, we have found more efficient
not to compare the Slater matrix to a common reference (typically, the Hartree-Fock determinant)
but instead to perform the Sherman-Morrison updates with respect to the
previously computed determinant $D_{i-1}^\sigma$.
To reduce the prefactor associated with this step
the list of determinants is sorted with a suitably chosen order
so that with high probability
successive determinants in the list differ only by one- or two-column substitution, thus
decreasing the average number of substitution performed.\\

$\bullet$ {\it Exploiting high-performance capabilities of present-day processors}.
This very practical aspect -- which is in general too much underestimated --
is far from being anecdotal since it allows us to gain important computational savings.
A number of important features include 
the use of vector fused-multiply add (FMA) 
instructions (that is, the calculation of \texttt{a=a+b*c} in one CPU cycle) for the innermost loops.
It is extremely 
efficient and should be systematically searched for. Using such instructions (present in general-purpose processors),
up to eight FMA per CPU cycle can be performed in double precision. While computing loops, overheads
are also very costly and should be reduced/eliminated. 
By taking care separately of the various parts of the loop
(peeling loop, scalar loop, vector loop, and tail loop) through size-specific and/or
hard-coded subroutines, a level of 100\% vectorized loops is reached in our code.
Another crucial point is to properly manage the data flow arriving to the processing unit.
As known, to be able to move data from the memory to the CPU with a sufficiently high data transfer to
keep the CPU busy is a major concern of modern calculations.  
Then, it is not only important to make maximum use of the low-latency cache memories to
store intermediate data but also to maximize prefetching allowing
the processor to anticipate the use of the right data and instructions in advance. 
To enhance prefetching the algorithm should allow the predictability 
of the data arrival in the CPU (that is, avoid random access as much as possible). 
It is this important aspect that has motivated us to use Sherman-Morrison updates,
despite the fact that a method like the Table method\cite{Clark_2011}
has formally a better scaling. Indeed, massive calculations of scalar products 
at the heart of repeated uses of SM updates are so ideally adapted to
present-day processors that very high performances can
be obtained.\\

$\bullet$ {\it Improved truncation scheme.} 
Instead of truncating the CI expansion according to the magnitude of the multideterminant 
coefficients as usual done, we propose instead to remove spin-specific determinants 
according to their total contribution to the norm
of the expansion. In this way, more determinants can be handled for a price corresponding to shorter
expansions.
To be more precise, we first observe that truncating the wavefunction according to the magnitude 
of coefficients has the effect of removing elements of the sparse matrix ${\bf C}$ 
of Eq.(\ref{eq:main}). A reduction of the computational cost occurs
only when a full line ($\uparrow$) or a full column ($\downarrow$) of ${\bf C}$ 
contains only zeroes, in that case
the determinant ${\bf D}_\sigma$ can be removed from the calculation.
Now, by expressing the norm of the wave function as
\begin{equation} 
   {\cal N} = \sum_{i=1}^{N_{\rm dets}^\uparrow} \sum_{j=1}^{N_{\rm dets}^\downarrow} C_{ij}^2 =  
    \sum_{i=1}^{N_{\rm dets}^\uparrow} {\cal N}_i^\uparrow = \sum_{j=1}^{N_{\rm dets}^\downarrow} {\cal N}_j^\downarrow. 
\label{eq:norm}
\end{equation}
it is possible to assign a contribution to the norm to each determinant. Then, all determinants
whose contribution to the norm is below some threshold
will be removed from the expansion.
This truncation scheme allows to eliminate the smallest number of coefficients needed to
obtain some computational gain. Moreover, the size-consistence property
of the wave function is expected to be approximately preserved by such
a truncation : when a $\sigma$-determinant is removed, it is
equivalent to removing the product of ${\bf D}_\sigma$ with 
all the ${\bar \sigma}$-determinants of the wave function. 

\section{Pseudopotentials for DMC using CIPSI}
\label{pseudo}
When using pseudopotentials a valence Hamiltonian is defined 
\begin{equation}
H_{val} = H_{\rm loc} + V_{\rm ECP}
\end{equation}
where $H_{\rm loc}$ is the local part describing the kinetic energy, the Coulombic repulsion and the local 
part of the effective core potentials (ECP).
\begin{equation}
H_{\rm loc} = -\frac{1}{2} \sum_i  \nabla_i^2  + \sum_{i,\alpha} v_{\rm loc}(r_{i\alpha}) + \sum_{i<j} 
\frac{1}{r_{ij}}
\end{equation}
and $V_{\rm ECP}$ the non-local part written as
\begin{equation}
V_{\rm ECP}= \sum_{i,\alpha} \sum_l v_l (r_{i \alpha}) \sum_{m=-l}^l  Y_{lm}(\Omega_{i \alpha}) \int
d{\Omega^\prime_{i \alpha}} Y^*_{lm} ( \Omega^\prime_{i \alpha})
\end{equation}
where $v_l$ is a radial pseudopotential, $Y_{lm}$ is the spherical harmonic, $\alpha$ 
labels pseudo-ions.

The action of a non-local operator being difficult to sample in DMC, 
$V_{\rm ECP}$ is ``localized'' by projecting 
it on the trial wavefunction. The localized form of the pseudo-potential is thus defined as
\begin{equation}
V^{\rm loc}_{\rm ECP}= \frac{V_{\rm ECP}\Psi_T}{\Psi_T}
\end{equation}
and we are led back to standard DMC simulations using only local operators 
at the price of introducing a new ``localization 
approximation''. This error is usually minimized by 
optimization of the trial wavefunction, see ref.\cite{casula}.
In practice, the necessity of numerically evaluating the localized potential is the main difference 
with standard DMC calculations. 

For each nucleus $\alpha$ and electron $i$, the two-dimensional angular integrals of the product of 
each $Y_{lm}$ and the trial wavefunction (all electrons fixed except the $i$th-electron moved 
over the sphere centered on nucleus $\alpha$ and of radius $r_{i \alpha}$) must be performed. 
By choosing the axes oriented 
such that the $i$th electron is on the $z$ axis, the contribution coming from the pair $(i,\alpha)$ 
is given by\cite{mitas_jcp_1991}
\begin{equation}
\sum_{i,\alpha} \sum_l \frac{2l+1}{4\pi} v_l (r_{i \alpha}) 
\int d{\Omega^\prime_{i \alpha}} P_l( cos \theta^\prime) \frac{ \psi_T ({\bf r}_1, ...{\bf r}^\prime_i,...,{\bf r}_N)} 
{\psi_T({\bf r}_1,..{\bf r}_i,.. {\bf r}_N)}
\end{equation}
where $P_l$ denotes a Legendre polynomial. Because of the Jastrow factor, 
the integrals involved cannot be computed analytically. 
The standard solution is to evaluate them numerically using some quadrature for the sphere.
Here, the CI form allows to perform the integration exactly, as already proposed 
some time ago.\cite{Hurley_1987,Hammond_1987}
Note that although no Jastrow prefactor is used here when localizing 
the pseudo-potential operator, such a prefactor can still be used for the DMC simulation itself.
A first advantage is that the calculation is significantly faster: in
practice, the computational cost is the same as evaluating the Laplacian of the wave function
and a gain proportional to the number of 
quadrature points is obtained. A second advantage is the possibility of a better
control of the localization error by increasing the number of determinants.

To illustrate these statements, we have chosen to calculate the atomization energy of the
C$_2$ molecule at the Hartree-Fock, CIPSI, DMC-HF and DMC-CIPSI levels with and without 
pseudopotentials. All-electron HF or CIPSI calculations have been performed with 
the cc-pVTZ basis set. To allow meaningful comparisons, $1s$ molecular orbitals have been kept 
frozen in all-electron CIPSI calculations. 
Pseudopotential calculations were done using the pseudopotentials of Burkatzki
\textit{et al.}\cite{burkatzki} with the corresponding VTZ basis set.
The electron-nucleus cusps of all the wave functions were
imposed,\cite{maCusp,Kussmann_cusp,per:cusp} and no Jastrow factor was used.
For the sake of comparison, the same time step ($5 \times 10^{-4}$ au) was used for all-electron and
pseudopotential calculations, although a much larger time step could have been
taken with pseudopotentials.

\begin{table}[ht]
\begin{ruledtabular}
\begin{tabular}{lccccc}
              &  \multicolumn{3}{c}{Energy} & \multicolumn{2}{c}{Number of determinants}  \\
              &  C (a.u.)     &   C$_2$ (a.u.)    &  AE (kcal/mol)  &  C  &   C$_2$     \\
\hline
\multicolumn{6}{l}{Hartree-Fock} \\
all-$e$    & -37.6867  &  -75.4015  &   17.6   & 1 & 1 \\
pseudo-  &  -5.3290  &   -10.6880 &   18.8   & 1 & 1 \\
\hline
\multicolumn{6}{l}{CIPSI} \\
all-$e$    & -37.7810 &  -75.7852  & 140.1  &  3796  &   $10^6$     \\
pseudo-  &  -5.4280 &   -11.0800 & 140.6  &  3882  &   $10^6$ \\
\hline
\multicolumn{6}{l}{DMC-HF} \\
all-$e$    & -37.8293(1) &  -75.8597(3) & 126.3(2) & 1 & 1  \\
pseudo-  &  -5.4167(1) &  -11.0362(3)  & 127.2(2) & 1 & 1  \\
\hline
\multicolumn{6}{l}{DMC-CIPSI, $\epsilon = 10^{-6}$} \\
all-$e$    & -37.8431(2) & -75.9166(2)  & 144.6(2)  & 3497     &   173553 \\
pseudo-  &  -5.4334(1) &  -11.0969(3)  & 144.3(2) &  3532    &   231991   \\
\hline
\multicolumn{3}{l}{Estimated exact AE \cite{Gingerich_1994,zpe_2007}} & 147$\pm$2 & & \\
\end{tabular}
\end{ruledtabular}
\caption{Comparison of all-electron (cc-pVTZ) and pseudopotential (BFD-VTZ) calculations of the atomization energy of C$_2$ with CIPSI wave functions. A threshold $\epsilon=10^{-6}$ was applied to the CIPSI wave functions as explained in text.}
\label{tab:pseudo}
\end{table}
The results presented in Table~\ref{tab:pseudo} show that all the atomization energies
obtained using pseudopotentials are in very good agreement with those obtained
with all-electron calculations at the same level of theory.
The DMC energies obtained with CIPSI trial wave functions are always below those
obtained with Hartree-Fock trial wave functions, and the error in the atomization energy
is reduced from 20 kcal/mol with HF nodes down to 3 kcal/mol with CIPSI nodes.

Calculations were performed on Intel Xeon E5-2680v3 processors.
Timings are given in Table~\ref{tab:timing}. For the carbon
atom the computational time needed for one walker to perform one complete Monte
Carlo step (all electrons moved) is the same with or without pseudopotentials.
For the C$_2$ molecule, the calculation is even faster with 
pseudopotentials: A factor of about $1.5$ is gained with respect to the 
all-electron calculation. This can be explained by the 
computational effort saved due to the reduced size of Slater matrices 
in the pseudopotential case (from $6\times6$ to $4\times4$) but, more importantly, by the fact 
that the additional cost related to the calculation of the contributions 
due to the pseudopotential is not enough important to reverse the situation.
In all-electron calculations, the variance is only slightly reduced when going from
the Hartree-Fock trial wave function to the CIPSI wave function (with frozen core). 
Indeed, the largest part of the fluctuations comes from the lack of correlation of the core electrons.
In the calculations involving pseudopotentials, the decrease of the variance is 
significant: a reduction by a factor of 2.4 and 3.2 is observed. 

From a more general perspective, comparisons between all-electron and pseudopotential calculations 
must take into account both the computational effort required in each case and the level 
of fluctuations resulting from 
the quality of the trial wavefunction. To quantify this, 
we have reported in the table the number of CPU hours
required to obtain an error bar of 1 kcal/mol.
Using pseudopotentials for the C$_2$ molecule, it is found that the reduction of the variance due
to the improvement of the wave function with the multideterminant expansion
almost compensates the cost of the computation due to the additional 230~000
determinants : the CPU time needed to obtain a desired accuracy is only $1.2\times$
more than the single determinant calculation. 
\begin{table}
\begin{ruledtabular}
\begin{tabular}{lcccccc}
       &\multicolumn{2}{c}{CPU time per DMC step} 
       &\multicolumn{2}{c}{CPU time to get a 1 kcal/mol} & \multicolumn{2}{c}{Variance} \\
       &\multicolumn{2}{c}{(milliseconds)} &\multicolumn{2}{c}{error (hours)} & \multicolumn{2}{c}{(a.u.)} \\
      \cline{2-3}	\cline{4-5}	\cline{6-7}
       &   all-$e$  &  pseudo-  &    all-$e$  &  pseudo-  &    all-$e$  &  pseudo-  \\
\hline
DMC-HF & &&&&&\\
C      &  0.0076  &  0.0078  &   1.54    &  1.18   & 7.858(3)  &  0.3471(2) \\
C$_2$     &  0.0286  &  0.0186  &  14.95     & 10.35  & 16.208(7) &  1.1372(6) \\
\hline
DMC-CIPSI  & &&&&& \\
C      &  0.193  &  0.201   &   5.61    &  0.70   & 7.620(8)  &  0.1084(4) \\
C$_2$  &  10.1   &  8.12    &  91.05    & 12.72  & 15.61(3)  &  0.460(1)  \\
\end{tabular}
\end{ruledtabular}
\caption{CPU time for one complete Monte Carlo step (one walker, all electrons moved), CPU time needed to reach an error on 1 kcal/mol, and variances associated with the HF and CIPSI trial wave functions (electron-nucleus cusp corrected).}
\label{tab:timing}
\end{table}


\section{Summary and some perspectives}
\label{conclu}

Let us first summarize the most important ideas and results presented in this work.\\
\\
i.) Selected Configuration Interaction approaches such as CIPSI are very efficient methods
for approaching the full CI limit with a number of determinants representing only a 
tiny fraction of the full determinantal space. This is so because only the most important 
determinants of the FCI expansion are perturbatively selected at each step of the iterative process. 
We note that the recent FCI-QMC method of Alavi {\it et al.}\cite{booth_jcp_2009,
cleland_jcp_2010} uses essentially the same idea, except that in CIPSI the selection is done 
deterministically instead of stochastically.
\\
\\
ii.) In constrast with exact FCI which becomes rapidly prohibitively expensive, CIPSI 
allows to treat larger systems, while maintaining results of near-Full CI quality.
The exact practical limits depend of course on the size of the basis set used, 
the number of active electrons, and also on the level of convergence asked for when
approaching the full CI limit. In this work, the CIPSI approach has been exemplified with
near-FCI quality all-electron calculations for the water molecule using a series of basis sets 
of increasing size up to the cc-pCV5Z basis set and for the whole set of 55 molecules and 9 atoms 
of the benchmark G1 set (cc-pVDZ basis set). In each case, the huge size of the FCI space forbids 
exact FCI calculations. CIPSI has been applied to larger systems, 
for example for calculating accurate total energies for the atoms of the $3d$ 
series,\cite{scemama_jcp_2014} 
and for obtaining near-FCI quality results for the CuCl$_2$ molecule 
(calculations including 63 electrons
and 25 active valence electrons).\cite{Caffarel_2014} 
Note that by using Effective Core Potentials as described in section \ref{pseudo} even 
larger systems can be treated.
\\
\\
iii.) We emphasize that the idea of selecting determinants is not limited 
to the entire space of determinants but can be used to make CI expansion to converge
in a subset of determinants chosen {\it a priori}. For example, efficient and accurate 
selected CASCI, CISD, or even MRCC\cite{mrcc_jcp_2016} calculations can be performed. Note that going beyond CASCI 
and implementing a selected CASSCF approach (CASCI with optimization of molecular orbitals) 
is also possible; this is let for further work. 
However, note that a stochastic version of CASSCF within FCI-QMC framework has already been 
implemented by Alavi {\it et al.}\cite{Thomas_jctc_2015}
\\
\\
iv.) CIPSI expansions can be used as determinantal part of the trial wavefunctions employed in 
DMC calculations. In others words, we propose to use selected CI 
nodes as approximation of the unknown exact nodes.
The basic motivation is 
that CI approaches provide a simple,
deterministic, and systematic way to build wavefunctions
of controllable quality. In a given one-particle basis set,
the wavefunction is improved by increasing the number of
determinants, up to the FCI limit. Then, by increasing
the basis set, the wavefunction can be further improved,
up to the CBS limit where the exact
solution of the continuous electronic Schr\"odinger equation is reached. CI
nodes, defined as the zeroes of the CI expansions, are also
expected to display such a systematic improvement.
\\
\\
v.) The main result giving substance to the use of selected CIPSI nodes is that 
in all applications realized so far the fixed-node error is found to decrease both 
as a function of the number of selected determinants and of the size of the basis set.
Mathematically speaking, such a result is far from being trivial.
In practice, such a property is particularly useful in terms of 
control of the fixed-node error. 
\\
\\
vi.) From a practical point of view, the price to pay is the need of 
considering much larger multideterminant expansions (from tens of thousands up to a few millions) 
than in standard DMC where compactness of the trial wavefunction is usually searched for.
Indeed, computing at each of Monte Carlo step 
the first and second derivatives of the trial wavefunction
(drift vector and local energy) is the hot spot of DMC.
However, efficient algorithms
have been proposed to perform such calculations\cite{Nukala_2009,Clark_2011,Weerasinghe_2014}. Here,
we have briefly summarized our recently introduced algorithm allowing to
compute $N$-determinant expansions issued from selected CI calculations
with a computational cost roughly proportional to $\sqrt{N}$ (with a small prefactor).
\\
\\
vii.) One key advantage of using CIPSI nodes is that their construction can be made 
fully automatic. Coefficients of the CI expansion are obtained in a simple and deterministic 
way by diagonalizing the Hamiltonian matrix and the solution is unique.
Furthermore, when approaching the FCI limit the resulting expansion becomes independent 
on the type of molecular orbitals used (canonical, natural, Kohn-Sham, see Figure 8 
of ref.\cite{Caffarel_2014}).
Another attractive feature is that the nodes built are reproducible and thus ``DMC models'' 
can be defined in the spirit of WFT or DFT {\it ab initio} approaches (HF/cc-pVnZ, MP2/6-31G, CCSD(T), DFT/B3LYP etc.)
Indeed, once the basis set has been specified, the nodes are unambiguously
defined at convergence of the DMC energy as a function of the number of selected determinants.
Furthermore, in this limit the nodal surfaces vary continuously as a function of the parameters of the 
Hamiltonian. A particularly important example is the possibility of obtaining regular 
potential energy surface (PES). This idea has been illustrated in a previous work on the 
potential energy curve of the F$_2$ molecule.\cite{giner_jcp_2015} Furthermore, 
it is also possible to reduce the ``non-parallelism'' error resulting from the use 
of a trial wavefunction of non-uniform quality across the PES. This can be done for example 
by using a variable number of selected determinants depending on the geometry
and chosen to lead to a constant second-order estimate of the remaining 
correlation energy (constant-PT2 approach,\cite{giner_jcp_2015} ). 
\\
\\
viii.) As in standard DMC approaches a Jastrow prefactor can be used to reduce statistical 
fluctuations. However, in contrast with what is usually done, we do not propose 
to re-optimize the determinantal CIPSI part 
in presence of this Jastrow term. The main reason for that is not to lose 
the advantages of using deterministically constructed nodal structures: Systematic 
improvement of nodes as a function of the number of determinants and of the size of 
the basis set, simplicity of construction of nodes and reproductibility, possibility of 
optimizing a very large number of small coefficients in the CI expansion (no noise limiting 
in practice the magnitude of optimizable coefficients), smooth evolution of nodes 
under variation of an external parameter (geometry, external field), etc.
\\
\\
ix.) The price to pay for not re-optimizing the determinantal part in the presence of a Jastrow
is that for small basis sets larger fixed-node errors are usually obtained. However, 
when increasing sufficiently the quality of basis set, it is no longer true as 
illustrated for example in the case of the oxygen atom,\cite{giner_2013} the water 
molecule,\cite{caffarel_jcp_2016} and the $3d$-transition 
metal atoms\cite{scemama_jcp_2014} for which benchmark total energies have been obtained.
\\
\\
x.) CIPSI wavefunctions are particularly attractive when using non-local Effective Core 
Potentials (ECP). Indeed, as already proposed some time ago,\cite{Hurley_1987,Hammond_1987}
CI expansions allow the analytical calculation of the action of the non-linear part 
of the pseudo-potential operator on the trial wavefunction.
In this way, the use of a numerical grid defined over the sphere is avoided 
and a gain in computational effort 
essentially proportional to the number of grid points is obtained.
Here, this idea has been illustrated in the case of the C$_2$ molecule. 
\\
\\
Finally, let us briefly mention a number of topics presently under investigation.
\\
\\
xi.) The slow part of the CI convergence is known to result from 
the absence of electron-electron cusp. In standard QMC approaches,
the short distance electron-electron behavior is introduced into the Jastrow prefactor 
and its impact on nodes is taken into account by optimization of the full
trial wavefunction.
Under re-optimization, molecular orbitals are changed and the distribution of 
multideterminant coefficients is modified with  
a re-inforcement of coefficients associated with chemically meaningful determinants and 
a reduction of the numerous small coefficients associated with the absence of cusp.
To keep the CIPSI as compact as possible and to eliminate this unphysical 
and uncoherent background of small coefficients 
a R12/F12 version of CIPSI is called for. We emphasize that 
such an analytical and deterministic construction of the R12/F12 expansion is necessary 
if we want to keep the advantages related to the deterministic 
construction of nodes.
\\
\\
xii.) To treat even larger systems, the increase of the number of determinants in the CIPSI 
expansion must be kept under control. 
Instead of targeting the near full CI limit, simpler models 
can be used in the spirit of what is 
done in MRCC approaches\cite{mrcc_jcp_2016}  
or by defining effective Hamiltonians in the reference space 
modelling the effect of the external space
(so-called internally decontracted approaches).
\\
\\
xiii.) Finally, it is clear that systematic studies on difficult systems of various types are 
needed to explore the potential and limits of the DMC-CIPSI approach.
\\
\\
{\it Acknowledgments.}
We would like to thank C. Angeli and P-F. Loos for their useful comments on the manuscript.
AS and MC thank the Agence Nationale pour la Recherche (ANR) for support through 
Grant No ANR 2011 BS08 004 01. 
This work was performed using HPC resources from CALMIP
(Toulouse) under allocation 2016-0510 and from GENCI-TGCC
(Grant 2016-08s015).


\begin{thebibliography}{96}
\bibitem [{not()}]{note2}%
  \BibitemOpen
  \href@noop {} {}\bibinfo {note} {An increase of the fixed-node energy may be
  sometimes observed at small number of determinants, (say, less than a few
  thousands), large basis sets, or when canonical orbitals are used. Up to now,
  this transient behavior has been found to systematically disappear when
  natural orbitals are used and/or larger expansion are
  considered.}\BibitemShut {Stop}%
\end{thebibliography}
%
\end{document}